\documentclass[]{interact}
\usepackage{epstopdf}% To incorporate .eps illustrations using PDFLaTeX, etc.
\usepackage[caption=false]{subfig}% Support for small, `sub' figures and tables
\usepackage[natbibapa]{apacite}

\usepackage{listings}
\theoremstyle{plain}% Theorem-like structures provided by amsthm.sty

\theoremstyle{definition}

\theoremstyle{remark}

\begin{document}

%%%%%%%%%%%%%%%%%%
% HPA Uitgezet %
%%%%%%%%%%%%%%%%%%.
% \articletype{ARTICLE TEMPLATE} % Specify the article type or omit as appropriate

\title{Improving students' code correctness and test completeness by informal specifications}

\author{
   \name{Arno Broeders\textsuperscript{a} and
        Ruud Hermans\textsuperscript{a} and
				Sylvia Stuurman\textsuperscript{b} and
				Lex Bijlsma\textsuperscript{b} and
				Harrie Passier\textsuperscript{b}\thanks{CONTACT Harrie Passier. Email: harrie.passier@ou.nl}
				}
\affil{\textsuperscript{a}Avans University of Applied Sciences, Professor Cobbenhagenlaan 13, 5037 DA Tilburg, The Netherlands; 
       \textsuperscript{b}Department Computer science, Open University, Valkenburgerweg 177, 6419 AT Heerlen, The Netherlands
			}
% \name{A.~N. Author\textsuperscript{a}\thanks{CONTACT A.~N. Author. Email: latex.helpdesk@tandf.co.uk} and John Smith\textsuperscript{b}}
%\affil{\textsuperscript{a}Taylor \& Francis, 4 Park Square, Milton Park, Abingdon, UK; 
%       \textsuperscript{b}Institut f\"{u}r Informatik, Albert-Ludwigs-Universit\"{a}t, Freiburg, Germany}
}

\maketitle

\begin{abstract}
The quality of software produced by students is often poor.
How to teach students to develop good quality software has long been a topic in computer science education and research. We must conclude that we still do not have a good answer to this question.

Specifications are necessary to determine the correctness of software, to develop error-free software and to write complete tests.
Several attempts have been made to teach students to write specifications before writing code.
So far, that has not proven to be very successful: Students do not like to write a specification and do not see the benefits of writing specifications.
 
In this paper we focus on the use of informal specifications.
Instead of teaching students how to write specifications, we teach them how to use informal specifications to develop correct software.
The results were surprising: the number of errors in software and the completeness of tests both improved considerably and, most importantly, students really appreciate the specifications.
We think that if students appreciate specification, we have a key to teach them how to specify and to appreciate its value.
\end{abstract}


\section{Introduction}

% https://www.tandfonline.com/action/journalInformation?show=aimsScope&journalCode=ncse20

The quality of software produced by students is often poor \citep{keuning2017code}.
Their software contains many errors and tests are missing or of low quality \citep{edwards2014student,lawende2021reproduction}.
Students use their own, incorrect, definition of correctness \citep{kolikant2005students,kolikant2008}.
For a long time, researchers and lecturers have been interested in how to improve teaching the value of good quality software and how to achieve that.
As far as we know, attempts to improve teaching have yielded little result so far.

It has often been argued that specifications or contracts \citep{parnas2011precise,passier2022,polikarpova2013good} are necessary to design and implement a correct program and to write good test cases. 
%Recently, we have come to understand how specifications can play a key role in learning how to design, implement, and test software.
%Programming is a complex task.
%To perform a complex task well, procedural knowledge is needed in addition to conceptual knowledge \citep{merrienboer2007}. Traditionally, conceptual knowledge gets a lot of attention in programming education.
%The different syntax constructions of a language are discussed and their application is shown by means of examples.
%Procedural knowledge, however, receives much less attention.
%Questions like `What steps are needed to develop a good program?', `What decisions must be made in the process?', and `How do I ensure the right tests?' are often taken implicitly or even not at all with all the consequences.
%Recent research shows that for the development of an OO class with an associated test, twelve main steps can be distinguished each with its own set of guidelines and rules of thumb.
%Specifications play a crucial role in this process \citep{passier2022}.
%We briefly describe this procedure in Section \ref{sec:complexity}. 
It is remarkable that in today's programming education much attention is paid to the syntax and the use of programming languages, but little or no attention is paid to what exactly a prospective program should do. If specifications are important, the question arises how students can best taught how to work with specifications. This paper attempts to provide insight into this question.

In the history of programming education, many attempts have been done to improve the quality of students' software by putting emphasis on specifications. We provide an overview of such attempts in Section \ref{sec:related}. Many of these proposals try to teach students to write a specification before they start programming. Teachers found that it was, indeed, possible (although difficult) to teach students to work like that for the duration of a course, but that students did not value the procedure, and abandoned it as soon as the course was finished. Students feel that to write an implementation is much easier than to write a specification, and, therefore, reverse the process in practice, i.e. first implement a function and than write the specification~\citep{hermansteaching2021,Brown-Requiring}.

Some of the attempts that were successful stress that specifications can only been taught to students with knowledge about mathematical logic, programming theory and specification theory, and thus cannot be taught to students who are beginning to learn to program~\citep{Back-2009}.

In this paper we focus on informal, possibly natural-language, method specifications in a style largely compatible with JML \citep{leavens1999jml}.  
In a recent attempt, first-year students were taught to create informal specifications. They were supported by procedural guidance \citep{hermansteaching2021}. The results were that students did not like to develop program specifications. They consider specification as a duplicate effort to coding. 

In practice, many teachers of first-year programming courses, especially those who start with object-oriented software, abandoned the idea to teach students to write a specification before starting to program.

%All these proposals assume a given formal specification used to reason about a program, to generate automatic tests, and/or to verify automatically the program's correctness using a tool. 
%Non of them tries to learn students to write specifications and/or to use them to implement the functionality or to develop test cases by hand. 

There are several disadvantages of abandoning specifications in first-year programming courses:

\begin{itemize}
	\item Students get used to write functions, methods or classes based on vague descriptions rather than on clear specifications.
	\item Students have no idea about correctness, because they only have those vague descriptions to base their ideas about correctness on.
	\item Students have nothing to base test cases on.
\end{itemize}

To overcome this dilemma (writing a specification first is considered too difficult, but leaving out specifications has too many and severe disadvantages), we propose an approach in which we give students specifications, based on which they write a program. We use informal but precise specifications to avoid the needed knowledge about logic notations. 

The questions we ask is whether this approach stimulates students to consider a specification as valuable, whether the quality of their programs increases, and whether their understanding of correctness improves:

\begin{enumerate}
	\item Does providing explicit specifications and procedural guidance in how to use them, help to improve the correctness of the code of students?
	\item Does providing explicit specifications and procedural guidance in how to convert them to test cases, help to improve the completeness of the test cases of students?
	% \item Does the understanding of correctness of students improve when we give them informal specifications?
	\item Do students see the value of specifications when they receive them to base test cases and implementations on them?
\end{enumerate}

To answer these questions, we provided each programming task with an informal but complete specification and taught students how to read a specification. Thereafter we taught them how to use it to implement code and to develop a corresponding test.

We show an example of an informal specification in Section \ref{sec:example}. Our research method is described in further detail in Section~\ref{sec:method}.
The results can be found in Section~\ref{sec:results}.
We discuss the implications of our results in Section~\ref{sec:discussion}.
Our answers to the research questions and how we will continue this research are described in Section~\ref{sec:conclusions}.
\section{Related work}
\label{sec:related}
Parnas has described in detail the role of software documentation, how it can be produced, and how it can be used \citep{parnas2011precise}. Specifications are important parts of software documentation. Among the benefits are correct implementation and more effective testing. 

In the past, many suggestions have been made to improve the quality of students' software by emphasizing specifications.
One such proposal is the \emph{Refinement calculus} proposed by Morgan \citep{morgan1990programming}, Morris \citep{morris1989laws} and others.
Here an initial specification undergoes stepwise decomposition into subtasks that are themselves either specified or implemented; the process ends when no unimplemented subtasks remain.
A disadvantage of this approach is that a great deal of pen-and-paper argumentation is necessary to produce even the simplest programs, which discourages its use as a feasible strategy in programming courses.

Extensive experience, on the other hand, exists with Dijkstra's \emph{Program derivation} method \citep{dijkstra1976discipline,dijkstra1988method,gries1981science}.
Here a given specification, consisting of a pre- and postcondition pair in first-order logic, is used as the starting point for a calculation with predicates, based on proof rules for the programming language constructs.
The main creative step in such a calculation is the choice of a loop invariant, for which several heuristics are available.
For at least a decade, first- and second-year students at Eindhoven University of Technology exclusively worked in this format \footnote{During this period author Bijlsma was one of the teachers involved} \citep{perrenet2005exploring}.
This resulted in programs in the \emph{guarded command} notation, designed on purpose to discourage running or testing the programs: the language contained no input or output statements and no procedure calls and the programs were inherently nondeterministic.
The underlying thought was that programs should possess guaranteed correctness through the method of their construction: programs and their correctness proofs were to be developed hand in hand, with the proof leading the way.
The experience with teaching this method showed that it was indeed possible to train students in applying this method, but that in general they found it too burdensome to follow its prescriptions in real-life programming tasks.
The Dafny system \citep{leino2014dafny} aims to take over some of the proof burden in this system through employment of an automated theorem prover.
Gegg-Harrison et al.\ \citep{gegg2003studying} employ the same semantics; however, they do not apply these to program derivation but to ex-posteriori correctness proofs.
Platzer \citep{platzer2013teaching} uses a similar approach in the context of cyber-physical systems, using the argument that people bet their lives on such systems, so they had better be correct.

Apart from their use in constructing the program itself, specifications are also useful for providing test cases.
Cheon and Leavens \citep{cheon2002simple} propose a testing system based on cooperation of JML \citep{leavens1999jml} and JUnit \citep{gulati2017java}, where the tester need only provide the test case input values.
No prediction of the expected result is required: the JML specification simply functions as a \emph{test oracle} by checking the state actually produced against the specification.
This only works if the specification is both formal and complete. 
However, as JML does not offer a full range of mathematical concepts, the expressiveness of purely formal JML specifications is limited; hence formality and completeness are not easily reconcilable.
Examples of this approach in action are given by du Bousquet et al.\ \citep{bousquet2004case} and Liu \citep{liu2007integrating}.

\emph{Design by Contract} is the specification style advocated by Meyer \citep{meyer2002design}.
In contrast to JML, this style does not focus on unit tests and does not require test cases to be specified in advance.
Instead, programs are adorned with inline assertions and there is no separate test class.
In the original view, these assertions are merely boolean expressions (in the Eiffel programming language \citep{meyer1997construction}, so without quantifiers or nontrivial mathematics, and necessarily formal.
These assertions are left in the code and are automatically tested at each run of the program.
Paige and Ostroff \citep{paige2004specification} consider its use as a way to get acquainted with formal methods.
Polikarpova et al.\ \citep{polikarpova2013good} propose to extend the expression language used in Design by Contract with references to simple mathematical structures such as sequences.
Experiments show that testing against such `strong specifications' detects twice as many bugs as standard contracts, with a reasonable overhead in terms of annotation burden and run-time performance while testing.
One other disadvantage of the restriction to executable assertions remains: it excludes information hiding, so Liskov's substitution principle \citep{liskov1994behavioral} does not apply \citep{Leino2002}.
Indeed, in Eiffel inheritance does not imply subtyping.

\emph{Procedural Guidance} \citep{passier2022} scaffolds the development of class and method design and  informal specifications. 
It distinguishes external design and specification (a.k.a. blackbox), internal design and specification (a.k.a. greybox), and implementation (a.k.a. whitebox). Furthermore, guidelines are given how to derive test cases from these specifications using the techniques equivalent classes and boundary testing. In the present paper, only the parts external design and specification, implementation, and corresponding tests are used. 

Liskov and Guttag develop classes and methods in a way resembling our proposal~\citep{LiskovGuttag2000,liskov1977}.
The starting point for distinguishing external and internal specifications goes back to the work of Tony Hoare~\citep{hoare1972}.
An informal notation is used to describe precondition/postconditions pairs. Differences are that exceptions are not described in a separate tag, but are included in the postcondition tag \textsc{effects} and where we use a \textsc{jml}-based notation, a custom notation has been chosen by them. 
A more fundamental difference is that they limit the \emph{relation} between model entities and implementation variables to be functional (from implementation to model) and use the term abstraction function as such in stead of representation relation in our approach. 
Finally, they do not describe a general process of how to develop specifications stepwise. 

Felleisen et al. describe an high level approach \citep{FelleisenFindler2001} consisting of six phases. Each phase is described in terms of a goal and activities. In the first phase, Data Analysis and Design, a data definition is formulated describing the interesting aspects of the objects mentioned in a problem statement. In the second phase, Contract, Purpose and Header, a contract is described naming the function and specifying the input and output data. The third phase, Examples, characterize input-output relationships via examples. In the fourth phase, Template, the body's outline is formulated. The fifth phase, Body, implements the body of the function. The last phase, Test, applies the function to the inputs of the examples and checks the outputs are as predicted. 
The approach asks for a description of the function's purpose on a high level. This description is not in terms of pre- and postconditions and no difference is made between `happy' and 'non-happy' behavior.   
Furthermore, because a functional language is taught, the approach doesn't make distinction between internal and external specifications.

\section{Example specification}\label{sec:example}

The following example shows a specfication that students receive as part of an exercise to implement the corresponding functionality and/or test cases. It is a specification of a method. 
Besides a general description of the method's functionality, the functionality is specified in terms of precondition/postcondition pairs. 
Using subspecifications, the JML-like specification distinguishes happy path behavior (everything is going well) and robustness (one or more parameters have forbidden values).    

\begin{lstlisting}[language=Java]
/**
 * @desc Calculates x to the power y
 * 
 * @sub Happy path {
 *   @requires x >= 0, y >= 0 and not (0^0);
 *   @ensures \result = x^y;
 * }
 *
 * @sub Zero {
 *   @requires x == 0 and y == 0;
 *   @ensures \result = 1;
 * }
 * 
 * @sub Negative {
 *   @requires x < 0 or y < 0;
 *   @signals IllegalArgumentException;
 * }
 *
 * @robust
 * @pure
 */
 public long power(int x, int y) throws IllegalArgumentException
\end{lstlisting}
   
% Besides for coding, the specifications are suited to specify test cases too.
% In doing so, we use equivalent classes and boundary values \citep{jorgensen2013software}. 

\section{Method}\label{sec:method}
An existing course Programming 2 at a university of applied sciences has been extended with two lectures on software quality and testing.
Both courses Programming 1 and 2 use the language Java. 

Students are taught how specifications should be read and how these specifications can be used when implementing code and designing tests to increase quality.
Each lecture consists of two hours of theory supplemented by a three hours supervised tutorial.

In the first lecture an introduction is given to the subject of the quality parameters correctness and completeness.
Students practise their feeling on correctness by manual testing precompiled class-files without source-code. These class-files contain intentional bugs.
To gain insight on correctness and ambiguity, students are challenged to find these bugs without extensive specifications, only a single fuzzy phrase that describes the working of the code is given. As an example: 'This method calculates the amount of time until a given birthday'.
Students execute the precompiled class-file with self-made parameters. The correctness is now derived from the comparison of the result of the output with students assumption. Through discussion afterwards, we let students experience that there is a lot of ambiguity in this given fuzzy specification.

After this first exercise we introduce extensive specifications as in the form described in section~\ref{sec:example} that have been partitioned using subspecifications and let the students find intentional bugs with special focus on edge cases. These specifications are stated unambiguous with semi-formal pre- and postconditions. The aim for students is to become aware that it is important to specify the specifications exactly to avoid ambiguity. A second goal is to let students experience that mostly several tests are needed to demonstrate the correctness of a software unit. Not only happy-path, but also the specified partitions with alternative paths should be tested.
The discussion with students afterwards should emphasize the importance of such detailed specifications.

In the second lecture there is focus on constructing external tests, implementing the functionality, and running the test, using JUnit as a test framework. 
During the lecture, several small exercises are discussed, in which the importance and the usability of explicit stated extensive specifications is made clear.
We teach students how to write code using a stripped-down version of the test-driven design methodology by using a procedural guide.

The following procedural steps were followed:
1) prepare the test suite and the project,
2) design and implement test cases for each given external (sub-)specification,
3) analyse, design and code the implementation,
4) run the test against the implementation.
Each of these steps is supported by detailed guidelines according to the procedure.

After the lectures, students complete a final assignment in teams of two.
Four methods are prescribed with complete specifications.
Each team chooses two of the four methods to be provided with test cases, but they implements all four methods.
Test cases may be added to the other two methods at free will. No extra credit is given if students test these extra methods.
Students take a screen-recording of their IDE with audio while they think aloud about the steps in the procedural guide.
With this setup, we can determine if students also create tests on their own and if there are differences between implementations with and without test cases.

All implementations are checked for correctness against a teacher test that is fully reviewed.
To check the correctness of the tests, each student submitted test runs against a correct and incorrect teacher implementation.
To check the completeness of the test sets, we will modify the correct teacher implementations that lend themselves to it by mutation testing.
Each mutation of the teacher implementation must result in the failure of at least one test. If this does not happen, the test set is incomplete.

After a few months, interviews and questionnaires are conducted to see whether students have adapted their coding-behavior and whether there is still sufficient knowledge and awareness regarding quality and testing.

\section{Results}\label{sec:results}

\subsection{Behavior}
We are interested in observing any differences in coding behavior between students who use a test-first approach based on external specifications together with procedural guidance, and students who do not use testing but have access to the specifications.

As to the perceived usefulness of testing, it was noted that five of the twenty-five student couples designed test suites for all four implementations while only two were required. Their reason for doing so was that, if the skills are present to make test suites, creating a test suite takes no extra time beyond trying manually to see if the implementation works, but creates more confidence in the program's correctness.

As to the interaction of specification and testing, conversion of specifications to test
suites went reasonably well. From the preconditions of the specifications, the test values are chosen and the oracle is derived from the post-condition. The procedural guide
was broadly followed. This does not mean that the guide needs to be consulted all the time: in the interviews, students indicate that the procedural guide is useful when learning the steps, but after performing this a few times during exercises, it goes automatically because these are fairly easy steps to master. One aspect is not going well: students quickly fall into making one test case per subcontract and forget to test the boundaries.

When creating the implementations that do not need test suites, parts of the specifications are still copied and used almost one-to-one as implementation code. Students find the specifications very useful for this, even if they do not write tests.

\subsection{Correctness}
Next we investigate what improvement can be seen in the code correctness when using a test-first approach together with explicit informal specifications and the procedural guide. The observations were made in the context of four separate exercises. The exercises themselves, as well as more extensive numerical findings, can be found in Broeders' thesis \citep{Broeders}.

\subsubsection{Exercise 1}
This was a very simple exercise that could easily be implemented with a one-liner. For this assignment, there are fourteen student couples (56\%) who submitted the implementation with a test suite. The remaining eleven student couples (44\%) only made the implementation of this method.

Of the fourteen student couples who made a test set, twelve student couples turned out to have made a completely correct implementation. This is 86\% of the group who wrote a test set for this method. Of the eleven student couples that chose not to make a test set, eight student couples turned out to have made a completely correct implementation. This is 73\% of the non-testing group. In order not to let students with really bad programming skills obscure the results, we also take the median of the correctness. Both groups score here the expected 100\% correctness for the implementation.

Of the fourteen test sets made, ten (71\%) sets test correctly. Surprisingly,
there are two student couples with good implementation but failing tests. When we look
at the recorded `thinking aloud' session, we see that these groups do pre-create test sets, they do make the implementation, but forget to test the implementation against their test set. We surmise from this behavior that testing with these students is not seen as a tool to ensure the correctness of the implementation, but merely as a task from school to be completed.

To assess the completeness of the test sets, we use mutation testing on the teacher implementation.
Any mutation made to the teacher implementation must fail at least one
test from the submitted test set. We see that eight of the fourteen test sets (57\%) are complete.
On average we see that 81\% of the mutations are detected, the median on-mutation
detection for this method is 100\%.
\subsubsection{Exercise 2}
For this assignment, there are sixteen student couples (64\%) who submitted the implementation
with a test suite. The remaining nine student couples (36\%) only made the
implementation of this method.
Of the sixteen student couples who also made a
test set, fourteen student couples of the testing group (88\%) turned out to have made a completely correct implementation.
Of the nine student couples that chose not to make a test set, five student couples of the non testing group (56\%) turned out to have made a
completely correct implementation.

Of the sixteen test sets made, thirteen (81\%) sets test correctly. The average test completeness over all submitted tests is 72\%, the median is 76\%.

\subsubsection{Exercise 3}
For this exercise, only one pair of students produced a completely correct implementation. Most of the others became confused because of an incorrect application of the Java \lstinline{String.split}
method and its interaction with regular expressions, a complication that had not been foreseen when designing the exercise.

As the percentage of correct implementations does not permit comparison between the two student groups, we turn to a measure for the number of failing test cases among the test cases provided by the teacher.  
In the group of students who also made test sets, an average of 28\% (median 21\%) of the teacher tests fail.
Of the group of students who did not write a test set for this method, on average 56\% (median 68\%) of the
teacher tests fail.

\subsubsection{Exercise 4}
This exercise turned out to be much harder: no single student couple has managed to implement this method correctly. 
Striking about the elaboration of this method is the fact that lots of students have not yet mastered the skills of throwing exceptions and testing these exceptions, even though this was already discussed in class weeks ago. In the `thinking aloud' session, we see them struggle to get
the exceptions working.

We do see major differences in failing teacher's test cases between the elaborations of students who have written tests in advance and students who have only made the implementation. To wit: of the teacher
test set, which consists of 20 test cases, an average of 28\% (median 10\%) of the tests fail in students who have written tests in advance, compared to 37\% (median 25\%) of the test cases in student couples who have not written any tests for their implementation.

\subsection{The role of specifications}
In the questionnaire we asked students whether the specifications helped with
completing the assignment. We received the answers of twenty students. Here 50\% answered in the affirmative for both coding and test construction, 25\% for test construction alone. Moreover, students explain that specifications
help to provide clarity about how the implementation should work, and they provides guidance in
determining the boundaries to be tested. It was even so clear that the specifications could
be mapped one-on-one to the tests and implementation. The 50\% of students who did not find
specifications helpful indicate that they find it a strange concept to `code’ the specifications
and then code exactly the same in the tests and after one time again in the
implementation. They indicate that they do not see the added value above an informal
comment in descriptive text prior to the method.

It should be noted that in these simple exercises, the correspondence between specification, code and tests is just a one-to-one translation to a different format. The role of specifications as communication between separate teams has not yet been discussed, and neither has their role in preserving architectural decisions during program evolution. None of the exercises featured specifications where a choice between several equally natural solutions had to be made.

Next, the questionnaire asked whether students would be willing to compose
specifications themselves from a problem statement in a natural language.
The answer to this question shows that less than half, with or without
the help of procedural guidance, would be willing to do this. Six students answered that they
only do this if it is actually graded, eight students indicate that they would write specifications, only in
a less formal way, eight answers indicate that they do not want to do it because it takes too
much time. Composing semi-formal specifications is therefore considered as cumbersome
and time-consuming.

\subsection{Analysis}
The specifications give students a guideline in designing the tests, it gives them an indication of which tests are needed and also helps students to code the implementation. This convenience is recognized by many students who participated in the interviews and questionnaire. We also saw in the `thinking aloud' sessions that students really use the specifications as a guideline for designing the tests and also for coding the implementation.

A test-first approach based on explicit specifications does help to improve code correctness among students. In comparison to other research, for example \citep{edwards2014student,lawende2021reproduction} the percentage of errors found is remarkable (80\%).    
The analysis of the assignments proves that the correctness of the implementations for which test suites have been written score significantly higher than implementations for which no tests have been written. From the `thinking aloud' sessions, we see that students also make extensive use of the specifications for implementations. And even if they do not create test suites, this still has a positive effect on code correctness.

Procedural guidance has an indirect positive effect on the code correctness of the implementations. We have rarely seen students consult the guide physically in the `thinking aloud' sessions, but it appears from these recordings and the submitted elaborations of the assignments that the steps have been completed successfully for the most part.

Despite the benefit students get from the specifications, students themselves indicate that they do not want to invest the time in drawing up the specifications so explicitly. Some students indicate that they may want to do this in an informal commentary style, but not in such a formal JML style as we have used in the given specifications. This could be due to the fact that students think they are only drawing up the specifications for themselves. Perhaps the teachers should put the drafting of explicit specifications into a practical situation where one student draws up explicit specifications from the requirements and shares them as a way of communication with another student who codes the tests or implementation. Also, examples should be provided with more algorithmic complexity, so that the specification is easier to read and understand than the code. In this way, explicit specification is experienced as more useful and we may get more willingness among the students.

\section{Discussion}\label{sec:discussion}
Trying to teach students to formulate specifications before they start coding has a long history. The experience of programming teachers always showed that students start to code, and then, after having completed their code, create specifications to please their teachers \citep{Brown-Requiring}.
However, the reason why teachers want students to write specifications beforehand is because it helps them to have a better understanding of the problem and to write better code. Obviously, forcing them to write specifications (whether these are made using a formal language or written in natural language) does not give students this insight.

One of the reasons for this resistance of students against writing specifications might be their vague understanding of the concept of correctness \citep{kolikant2005students}. When you do not understand the meaning of correctness, it is impossible to write a precise and complete specification. When we ask students to write specifications before they really understand the concept of correctness,  we violate against the scaffolding principle. All in all, teaching students to write specifications before writing code has proved to be an ineffective way to teach them what correct programs are and it is understandable why this is the case \citep{hermansteaching2021}. 

Our result, that students really appreciate the specifications that we provide in the form of contracts in natural language, and that they actually write better code, made us think about the practice of teaching students to program.

As an alternative to asking students to write specifications, many teachers have chosen to start with programming. Specifications are considered an advanced topic, something that can only be learned after having mastered how to program. However, when we teach programming like that, we often give them examples, exercises and assignments with a vague description of the problem. 

An example from a book teaching Python is the following: 'Write a function to compute $x$ to the power of $y$'. The solution, according to the book, is:

\begin{lstlisting}[language=Python]
	def power(x, y)
	  r = 1
	  while y > 0:
	    r = r * x
	    y = y - 1
	  return r
\end{lstlisting}
Using such examples, we teach students to make implicit assumptions and use those assumptions in the code.
Here, for instance, students are expected to assume:
\begin{itemize}
	\item $y$ is an integer
	\item when $y$ is negative, the result will be 1
	\item $x$ is a numeral
\end{itemize}
Exactly those kind of assumptions are found in the way that students implement and test their code, as we found out~\citep{Bijlsma-2021}.

Teaching students programming using vague problem descriptions, implicitly teaches them that making (optimistic) assumptions about input is 'good'. Implicitly, by giving vague problem descriptions, we teach them wrong ideas about correctness: exactly those ideas that are observed in students \citep{kolikant2005students}. The specification of the exercise used about the ATM algorithm (calculate the numbers of \$20 and \$50 bills the ATM should give for a given amount) is incomplete \citep{kolikant2004establishing}.
It was clear that amounts like \$10 and \$30 are special cases.
However, what the algorithm has to ensure in these cases was open.
As a result, several interpretations, differing in robustness, of correctness are in fact possible.     

Our approach avoids the pitfalls of both alternatives. We provide scaffolding, by giving students specifications instead of asking them to write them themselves. The specifications are easily readable, in the form of contracts. We also avoid implicitly teaching them to make assumptions. Instead, we create the expectation that, to write code, you need an exact specification. 
On top of that, a specification makes it easier to create code. 

We see the following advantages of our approach:

\begin{itemize}
	\item Students implicitly learn that a problem should be well defined before you start to code.
	\item Students implicitly learn what correctness is.
	\item Coding becomes easier when given good specifications. Our approach makes it easier to learn to program.
	\item The specifications offer a good base to create more complete test cases.
\end{itemize}

There is only one disadvantage that we can think of. We teachers have to put more effort in creating examples, exercises and assignments. In fact, we have to do what we asked students to do for a long time (write a specification before starting to code).

%- Discussie / inzicht  (***Sylvia)
%
%* o.a. Kolikant bespreken
%
%* !!de schommels!! (zonder contract kan correctheid niet beoordeeld worden!) Studenten maken eigen aannames (impliciet wordt alleen aan happy path gedacht) die gecontroleerd moeten worden (en dat kan niet tijdens een tentamen).
%
%* Nieuwe didactiek: vanaf begin leren werken en werken met contracten (leren dan al welke aannames gamaakt moeten worden) en op zeker moment PG om zelf contracten op te stellen. Dit verlaagt ook sterk de complexiteit van het programmeren (alleen inzet van syntax/algoritmiek om functionaliteit gedefinieerd in contract te bedenken en te implementeren.
%
%* Formele notatie is niet nodig! Zie PG
%
%* Contract schrijven versus code schrijven versus testen schrijven

\section{Conclusions and future work}\label{sec:conclusions}

All our research questions can be answered positively.
If explicit specifications and procedural guidance are provided to students, then the correctness of students' code  improves significantly (research question 1) as well as the completeness  of the test cases of students (research question 2). Students experience the value of specifications, but the examples given so far have not convinced them of the added value of writing specifications themselves.

Of course, we have to find out whether our approach helps students to write specifications themselves in a later phase. 
We hope that they have learned, by the many examples of specifications, to see the gaps in the problem description.
They should be able to ask for more clarification when they are given a vague problem description or that they are able to draw up a complete specification in case of an incomplete specification for simple cases.

We see two directions for further research. 
First, we wonder how we can better convince students of the added value of drawing up and using specification. 
We will research this by offering more complex, real world assignments. For example, exercises: 
\begin{itemize}
	\item where one student draws up explicit specifications were another student codes the tests and/or implementation,
	\item with relative easy to read specifications, but ask for a complex algorithm to implement,
	\item in which a complex algorithm must be reviewed with and without accompanying specifications,
	\item in which parts of a project have to be refactored, with and without specifications. 
\end{itemize}
After students really understand the benefits of specifications, we can teach them how to construct these. 

Second, in this study we see that the correctness of students' programs increases and that students appreciate the value of specifications. To what extent, however, students also have a better understanding of correctness after working with specifications is still unclear \citep{kolikant2005students}. We will also research the misconceptions that are involved, like the assumption that only one test case is needed per subcontract \citep{Eldh-Analysis}.

% HPA: TODO %%%%%%%%%%%%%%%%%%%%%%%%%%%
%\section*{Acknowledgements TODO}
%\section*{Funding}
%%%%%%%%%%%%%%%%%%%%%%%%%%%%%%%%%%%%%%%

%\section*{Funding}
%
%An unnumbered section, e.g.\ \verb"\section*{Funding}", may be used for grant details, etc.\ if required and included \emph{in the non-anonymous version} before any Notes or References.
%
%
%
%
%
%\section{References}
%
%\subsection{References cited in the text}
%
%References should be cited in accordance with \citeauthor{APA10} (APA) style, i.e.\ in alphabetical order separated by semicolons, e.g.\ `\citep{Ban77,Pia88,VL07}' or `\ldots see Smith (1985, p.~75)'. If there are two or more authors with the same surname, use the first author's initials with the surnames, e.g.\ `\citep{Lig08,Lig06}'. If there are three to five authors, list all the authors in the first citation, e.g.\ `\citep{GSSM91}'. In subsequent citations, use only the first author's surname followed by et al., e.g.\ `\citep{GSSM91}'. For six or more authors, cite the first author's name followed by et al. For two or more sources by the same author(s) in the same year, use lower-case letters (a,~b,~c, \ldots) with the year to order the entries in the reference list and use these lower-case letters with the year in the in-text citations, e.g.\ `(Green, 1981a,b)'. For further details on this reference style, see the Instructions for Authors on the Taylor \& Francis website.
%
%

\section*{Funding}
This research is co-funded by the Erasmus+ Programme of the European Union. 

\bibliographystyle{apacite}
\bibliography{contracts}

\begin{thebibliography}{}

\bibitem [\protect \citeauthoryear {%
Back%
}{%
Back%
}{%
{\protect \APACyear {2009}}%
}]{%
Back-2009}
\APACinsertmetastar {%
Back-2009}%
\begin{APACrefauthors}%
Back, R\BPBI J.%
\end{APACrefauthors}%
\unskip\
\newblock
\APACrefYearMonthDay{2009}{}{}.
\newblock
{\BBOQ}\APACrefatitle {Invariant based programming: basic approach and teaching
  experiences} {Invariant based programming: basic approach and teaching
  experiences}.{\BBCQ}
\newblock
\APACjournalVolNumPages{Formal Aspects of Computing}{21}{3}{227--244}.
\PrintBackRefs{\CurrentBib}

\bibitem [\protect \citeauthoryear {%
Bijlsma%
, Doorn%
, Passier%
, Pootjes%
\BCBL {}\ \BBA {} Stuurman%
}{%
Bijlsma%
\ \protect \BOthers {.}}{%
{\protect \APACyear {2021}}%
}]{%
Bijlsma-2021}
\APACinsertmetastar {%
Bijlsma-2021}%
\begin{APACrefauthors}%
Bijlsma, L.%
, Doorn, N.%
, Passier, H.%
, Pootjes, H.%
\BCBL {}\ \BBA {} Stuurman, S.%
\end{APACrefauthors}%
\unskip\
\newblock
\APACrefYearMonthDay{2021}{}{}.
\newblock
{\BBOQ}\APACrefatitle {How do students test software units?} {How do students
  test software units?}{\BBCQ}
\newblock
\BIn{} \APACrefbtitle {2021 IEEE/ACM 43rd International Conference on Software
  Engineering: Software Engineering Education and Training (ICSE-SEET)} {2021
  ieee/acm 43rd international conference on software engineering: Software
  engineering education and training (icse-seet)}\ (\BPGS\ 189--198).
\PrintBackRefs{\CurrentBib}

\bibitem [\protect \citeauthoryear {%
Broeders%
}{%
Broeders%
}{%
{\protect \APACyear {2021}}%
}]{%
Broeders}
\APACinsertmetastar {%
Broeders}%
\begin{APACrefauthors}%
Broeders, A.%
\end{APACrefauthors}%
\unskip\
\newblock
\APACrefYear{2021}.
\unskip\
\newblock
\APACrefbtitle {How do explicit specifications, a test-first approach, and
  procedural guidance to convert specifications to tests, help to improve code
  quality among students} {How do explicit specifications, a test-first
  approach, and procedural guidance to convert specifications to tests, help to
  improve code quality among students}\ \APACtypeAddressSchool {\BUMTh}{}{}.
\unskip\
\newblock
\APACaddressSchool {}{Open Universiteit}.
\unskip\
\newblock
\APACrefnote{Online at
  https://research.ou.nl/en/studentTheses/how-do-explicit-specifications-a-test-first-approach-and-procedur}
\PrintBackRefs{\CurrentBib}

\bibitem [\protect \citeauthoryear {%
Brown%
}{%
Brown%
}{%
{\protect \APACyear {1988}}%
}]{%
Brown-Requiring}
\APACinsertmetastar {%
Brown-Requiring}%
\begin{APACrefauthors}%
Brown, D\BPBI A.%
\end{APACrefauthors}%
\unskip\
\newblock
\APACrefYearMonthDay{1988}{}{}.
\newblock
{\BBOQ}\APACrefatitle {Requiring CS1 students to write requirements
  specifications: a rationale, implementation suggestions, and a case study}
  {Requiring cs1 students to write requirements specifications: a rationale,
  implementation suggestions, and a case study}.{\BBCQ}
\newblock
\APACjournalVolNumPages{SIGCSE bulletin}{20}{1}{13-16}.
\PrintBackRefs{\CurrentBib}

\bibitem [\protect \citeauthoryear {%
Cheon%
\ \BBA {} Leavens%
}{%
Cheon%
\ \BBA {} Leavens%
}{%
{\protect \APACyear {2002}}%
}]{%
cheon2002simple}
\APACinsertmetastar {%
cheon2002simple}%
\begin{APACrefauthors}%
Cheon, Y.%
\BCBT {}\ \BBA {} Leavens, G\BPBI T.%
\end{APACrefauthors}%
\unskip\
\newblock
\APACrefYearMonthDay{2002}{}{}.
\newblock
{\BBOQ}\APACrefatitle {A simple and practical approach to unit testing: The
  {JML} and {JUnit} way} {A simple and practical approach to unit testing: The
  {JML} and {JUnit} way}.{\BBCQ}
\newblock
\BIn{} \APACrefbtitle {European Conference on Object-Oriented Programming}
  {European conference on object-oriented programming}\ (\BPGS\ 231--255).
\PrintBackRefs{\CurrentBib}

\bibitem [\protect \citeauthoryear {%
Dijkstra%
}{%
Dijkstra%
}{%
{\protect \APACyear {1976}}%
}]{%
dijkstra1976discipline}
\APACinsertmetastar {%
dijkstra1976discipline}%
\begin{APACrefauthors}%
Dijkstra, E\BPBI W.%
\end{APACrefauthors}%
\unskip\
\newblock
\APACrefYear{1976}.
\newblock
\APACrefbtitle {A Discipline of Programming} {A discipline of programming}.
\newblock
\APACaddressPublisher{}{Prentice-Hall}.
\PrintBackRefs{\CurrentBib}

\bibitem [\protect \citeauthoryear {%
Dijkstra%
\ \BBA {} Feijen%
}{%
Dijkstra%
\ \BBA {} Feijen%
}{%
{\protect \APACyear {1988}}%
}]{%
dijkstra1988method}
\APACinsertmetastar {%
dijkstra1988method}%
\begin{APACrefauthors}%
Dijkstra, E\BPBI W.%
\BCBT {}\ \BBA {} Feijen, W\BPBI H.%
\end{APACrefauthors}%
\unskip\
\newblock
\APACrefYear{1988}.
\newblock
\APACrefbtitle {A Method of Programming} {A method of programming}.
\newblock
\APACaddressPublisher{}{Addison-Wesley}.
\PrintBackRefs{\CurrentBib}

\bibitem [\protect \citeauthoryear {%
Du~Bousquet%
, Ledru%
, Maury%
, Oriat%
\BCBL {}\ \BBA {} Lanet%
}{%
Du~Bousquet%
\ \protect \BOthers {.}}{%
{\protect \APACyear {2004}}%
}]{%
bousquet2004case}
\APACinsertmetastar {%
bousquet2004case}%
\begin{APACrefauthors}%
Du~Bousquet, L.%
, Ledru, Y.%
, Maury, O.%
, Oriat, C.%
\BCBL {}\ \BBA {} Lanet, J\BHBI L.%
\end{APACrefauthors}%
\unskip\
\newblock
\APACrefYearMonthDay{2004}{}{}.
\newblock
{\BBOQ}\APACrefatitle {A case study in {JML}-based software validation} {A case
  study in {JML}-based software validation}.{\BBCQ}
\newblock
\BIn{} \APACrefbtitle {Proceedings. 19th International Conference on Automated
  Software Engineering, 2004.} {Proceedings. 19th international conference on
  automated software engineering, 2004.}\ (\BPGS\ 294--297).
\PrintBackRefs{\CurrentBib}

\bibitem [\protect \citeauthoryear {%
Edwards%
\ \BBA {} Shams%
}{%
Edwards%
\ \BBA {} Shams%
}{%
{\protect \APACyear {2014}}%
}]{%
edwards2014student}
\APACinsertmetastar {%
edwards2014student}%
\begin{APACrefauthors}%
Edwards, S\BPBI H.%
\BCBT {}\ \BBA {} Shams, Z.%
\end{APACrefauthors}%
\unskip\
\newblock
\APACrefYearMonthDay{2014}{}{}.
\newblock
{\BBOQ}\APACrefatitle {Do student programmers all tend to write the same
  software tests?} {Do student programmers all tend to write the same software
  tests?}{\BBCQ}
\newblock
\BIn{} \APACrefbtitle {Proceedings of the 2014 conference on Innovation \&
  technology in computer science education} {Proceedings of the 2014 conference
  on innovation \& technology in computer science education}\ (\BPGS\
  171--176).
\PrintBackRefs{\CurrentBib}

\bibitem [\protect \citeauthoryear {%
Eldh%
, Hansson%
\BCBL {}\ \BBA {} Punnekkat%
}{%
Eldh%
\ \protect \BOthers {.}}{%
{\protect \APACyear {2011}}%
}]{%
Eldh-Analysis}
\APACinsertmetastar {%
Eldh-Analysis}%
\begin{APACrefauthors}%
Eldh, S.%
, Hansson, H.%
\BCBL {}\ \BBA {} Punnekkat, S.%
\end{APACrefauthors}%
\unskip\
\newblock
\APACrefYearMonthDay{2011}{}{}.
\newblock
{\BBOQ}\APACrefatitle {Analysis of mistakes as a method to improve test case
  design} {Analysis of mistakes as a method to improve test case
  design}.{\BBCQ}
\newblock
\BIn{} \APACrefbtitle {2011 Fourth IEEE International Conference on Software
  Testing, Verification and Validation} {2011 fourth ieee international
  conference on software testing, verification and validation}\ (\BPGS\
  70--79).
\PrintBackRefs{\CurrentBib}

\bibitem [\protect \citeauthoryear {%
Felleisen%
, Findler%
, Flatt%
\BCBL {}\ \BBA {} Krishnamurthi%
}{%
Felleisen%
\ \protect \BOthers {.}}{%
{\protect \APACyear {2001}}%
}]{%
FelleisenFindler2001}
\APACinsertmetastar {%
FelleisenFindler2001}%
\begin{APACrefauthors}%
Felleisen, M.%
, Findler, R\BPBI B.%
, Flatt, M.%
\BCBL {}\ \BBA {} Krishnamurthi, S.%
\end{APACrefauthors}%
\unskip\
\newblock
\APACrefYear{2001}.
\newblock
\APACrefbtitle {{How to Design Programs. An Introduction to Computing and
  Programming}} {{How to Design Programs. An Introduction to Computing and
  Programming}}.
\newblock
\APACaddressPublisher{Cambridge, Massachusetts London, England}{The MIT Press}.
\newblock
\APACrefnote{MIT -- Massachusetts Institute of Technology -- Online version
  $3^{rd}$ edition 22.~September~2002:
  \url{http://www.htdp.org/2002-09-22/Book/curriculum.html} -- last visited
  $19^{th}$~June~2008}
\PrintBackRefs{\CurrentBib}

\bibitem [\protect \citeauthoryear {%
Gegg-Harrison%
, Bunce%
, Ganetzky%
, Olson%
\BCBL {}\ \BBA {} Wilson%
}{%
Gegg-Harrison%
\ \protect \BOthers {.}}{%
{\protect \APACyear {2003}}%
}]{%
gegg2003studying}
\APACinsertmetastar {%
gegg2003studying}%
\begin{APACrefauthors}%
Gegg-Harrison, T\BPBI S.%
, Bunce, G\BPBI R.%
, Ganetzky, R\BPBI D.%
, Olson, C\BPBI M.%
\BCBL {}\ \BBA {} Wilson, J\BPBI D.%
\end{APACrefauthors}%
\unskip\
\newblock
\APACrefYearMonthDay{2003}{}{}.
\newblock
{\BBOQ}\APACrefatitle {Studying program correctness by constructing contracts}
  {Studying program correctness by constructing contracts}.{\BBCQ}
\newblock
\BIn{} \APACrefbtitle {Proceedings of the 8th annual conference on Innovation
  and technology in computer science education} {Proceedings of the 8th annual
  conference on innovation and technology in computer science education}\
  (\BPGS\ 129--133).
\PrintBackRefs{\CurrentBib}

\bibitem [\protect \citeauthoryear {%
Gries%
}{%
Gries%
}{%
{\protect \APACyear {1981}}%
}]{%
gries1981science}
\APACinsertmetastar {%
gries1981science}%
\begin{APACrefauthors}%
Gries, D.%
\end{APACrefauthors}%
\unskip\
\newblock
\APACrefYear{1981}.
\newblock
\APACrefbtitle {The Science of Programming} {The science of programming}.
\newblock
\APACaddressPublisher{}{Springer}.
\PrintBackRefs{\CurrentBib}

\bibitem [\protect \citeauthoryear {%
Gulati%
\ \BBA {} Sharma%
}{%
Gulati%
\ \BBA {} Sharma%
}{%
{\protect \APACyear {2017}}%
}]{%
gulati2017java}
\APACinsertmetastar {%
gulati2017java}%
\begin{APACrefauthors}%
Gulati, S.%
\BCBT {}\ \BBA {} Sharma, R.%
\end{APACrefauthors}%
\unskip\
\newblock
\APACrefYearMonthDay{2017}{}{}.
\newblock
{\BBOQ}\APACrefatitle {Java Unit Testing with {JUnit} 5} {Java unit testing
  with {JUnit} 5}.{\BBCQ}
\newblock
\BIn{} \APACrefbtitle {Java Unit Testing with JUnit.} {Java unit testing with
  junit.}
\newblock
\APACaddressPublisher{}{Springer}.
\PrintBackRefs{\CurrentBib}

\bibitem [\protect \citeauthoryear {%
Hermans%
}{%
Hermans%
}{%
{\protect \APACyear {2021}}%
}]{%
hermansteaching2021}
\APACinsertmetastar {%
hermansteaching2021}%
\begin{APACrefauthors}%
Hermans, R.%
\end{APACrefauthors}%
\unskip\
\newblock
\APACrefYear{2021}.
\unskip\
\newblock
\APACrefbtitle {Teaching Specification Writing To Improve Students’ Code}
  {Teaching specification writing to improve students’ code}\
  \APACtypeAddressSchool {\BUMTh}{}{}.
\unskip\
\newblock
\APACaddressSchool {}{Open Universiteit}.
\unskip\
\newblock
\APACrefnote{Online at
  https://research.ou.nl/en/studentTheses/teaching-specification-writing-to-improve-students-code}
\PrintBackRefs{\CurrentBib}

\bibitem [\protect \citeauthoryear {%
Hoare%
}{%
Hoare%
}{%
{\protect \APACyear {1972}}%
}]{%
hoare1972}
\APACinsertmetastar {%
hoare1972}%
\begin{APACrefauthors}%
Hoare, C\BPBI A\BPBI R.%
\end{APACrefauthors}%
\unskip\
\newblock
\APACrefYearMonthDay{1972}{}{}.
\newblock
{\BBOQ}\APACrefatitle {Proof of correctness of data representations} {Proof of
  correctness of data representations}.{\BBCQ}
\newblock
\APACjournalVolNumPages{Acta informatica}{1}{4}{271-281}.
\PrintBackRefs{\CurrentBib}

\bibitem [\protect \citeauthoryear {%
Keuning%
, Heeren%
\BCBL {}\ \BBA {} Jeuring%
}{%
Keuning%
\ \protect \BOthers {.}}{%
{\protect \APACyear {2017}}%
}]{%
keuning2017code}
\APACinsertmetastar {%
keuning2017code}%
\begin{APACrefauthors}%
Keuning, H.%
, Heeren, B.%
\BCBL {}\ \BBA {} Jeuring, J.%
\end{APACrefauthors}%
\unskip\
\newblock
\APACrefYearMonthDay{2017}{}{}.
\newblock
{\BBOQ}\APACrefatitle {Code quality issues in student programs} {Code quality
  issues in student programs}.{\BBCQ}
\newblock
\BIn{} \APACrefbtitle {Proceedings of the 2017 ACM Conference on Innovation and
  Technology in Computer Science Education} {Proceedings of the 2017 acm
  conference on innovation and technology in computer science education}\
  (\BPGS\ 110--115).
\PrintBackRefs{\CurrentBib}

\bibitem [\protect \citeauthoryear {%
Kolikant%
}{%
Kolikant%
}{%
{\protect \APACyear {2005}}%
}]{%
kolikant2005students}
\APACinsertmetastar {%
kolikant2005students}%
\begin{APACrefauthors}%
Kolikant, Y\BPBI B\BHBI D.%
\end{APACrefauthors}%
\unskip\
\newblock
\APACrefYearMonthDay{2005}{}{}.
\newblock
{\BBOQ}\APACrefatitle {Students' alternative standards for correctness}
  {Students' alternative standards for correctness}.{\BBCQ}
\newblock
\BIn{} \APACrefbtitle {Proceedings of the first international workshop on
  Computing education research} {Proceedings of the first international
  workshop on computing education research}\ (\BPGS\ 37--43).
\PrintBackRefs{\CurrentBib}

\bibitem [\protect \citeauthoryear {%
Kolikant%
\ \BBA {} Mussai%
}{%
Kolikant%
\ \BBA {} Mussai%
}{%
{\protect \APACyear {2008}}%
}]{%
kolikant2008}
\APACinsertmetastar {%
kolikant2008}%
\begin{APACrefauthors}%
Kolikant, Y\BPBI B\BHBI D.%
\BCBT {}\ \BBA {} Mussai, M.%
\end{APACrefauthors}%
\unskip\
\newblock
\APACrefYearMonthDay{2008}{}{}.
\newblock
{\BBOQ}\APACrefatitle {`{S}o my program doesn’t run!' {D}efinition, origins,
  and practical expressions of students’ (mis)conceptions of correctness}
  {`{S}o my program doesn’t run!' {D}efinition, origins, and practical
  expressions of students’ (mis)conceptions of correctness}.{\BBCQ}
\newblock
\APACjournalVolNumPages{Computer Science Education}{18}{2}{135-151}.
\newblock
\begin{APACrefURL} \url{https://doi.org/10.1080/08993400802156400}
  \end{APACrefURL}
\newblock
\begin{APACrefDOI} \doi{10.1080/08993400802156400} \end{APACrefDOI}
\PrintBackRefs{\CurrentBib}

\bibitem [\protect \citeauthoryear {%
Kolikant%
\ \BBA {} Pollack%
}{%
Kolikant%
\ \BBA {} Pollack%
}{%
{\protect \APACyear {2004}}%
}]{%
kolikant2004establishing}
\APACinsertmetastar {%
kolikant2004establishing}%
\begin{APACrefauthors}%
Kolikant, Y\BPBI B\BHBI D.%
\BCBT {}\ \BBA {} Pollack, S.%
\end{APACrefauthors}%
\unskip\
\newblock
\APACrefYearMonthDay{2004}{}{}.
\newblock
{\BBOQ}\APACrefatitle {Establishing computer science professional norms among
  high-school students} {Establishing computer science professional norms among
  high-school students}.{\BBCQ}
\newblock
\APACjournalVolNumPages{Computer Science Education}{14}{1}{21--35}.
\PrintBackRefs{\CurrentBib}

\bibitem [\protect \citeauthoryear {%
Lawende%
, Passier%
\BCBL {}\ \BBA {} Alp{\'a}r%
}{%
Lawende%
\ \protect \BOthers {.}}{%
{\protect \APACyear {2021}}%
}]{%
lawende2021reproduction}
\APACinsertmetastar {%
lawende2021reproduction}%
\begin{APACrefauthors}%
Lawende, M.%
, Passier, H.%
\BCBL {}\ \BBA {} Alp{\'a}r, G.%
\end{APACrefauthors}%
\unskip\
\newblock
\APACrefYearMonthDay{2021}{}{}.
\newblock
{\BBOQ}\APACrefatitle {Reproduction for Insight: Towards Better Understanding
  the Quality of Students Tests} {Reproduction for insight: Towards better
  understanding the quality of students tests}.{\BBCQ}
\newblock
\BIn{} \APACrefbtitle {Proceedings of the 26th ACM Conference on Innovation and
  Technology in Computer Science Education V. 1} {Proceedings of the 26th acm
  conference on innovation and technology in computer science education v. 1}\
  (\BPGS\ 213--219).
\PrintBackRefs{\CurrentBib}

\bibitem [\protect \citeauthoryear {%
Leavens%
, Baker%
\BCBL {}\ \BBA {} Ruby%
}{%
Leavens%
\ \protect \BOthers {.}}{%
{\protect \APACyear {1999}}%
}]{%
leavens1999jml}
\APACinsertmetastar {%
leavens1999jml}%
\begin{APACrefauthors}%
Leavens, G\BPBI T.%
, Baker, A\BPBI L.%
\BCBL {}\ \BBA {} Ruby, C.%
\end{APACrefauthors}%
\unskip\
\newblock
\APACrefYearMonthDay{1999}{}{}.
\newblock
{\BBOQ}\APACrefatitle {JML: A notation for detailed design} {Jml: A notation
  for detailed design}.{\BBCQ}
\newblock
\BIn{} \APACrefbtitle {Behavioral Specifications of Businesses and Systems}
  {Behavioral specifications of businesses and systems}\ (\BPGS\ 175--188).
\newblock
\APACaddressPublisher{}{Springer}.
\PrintBackRefs{\CurrentBib}

\bibitem [\protect \citeauthoryear {%
Leino%
\ \BBA {} Nelson%
}{%
Leino%
\ \BBA {} Nelson%
}{%
{\protect \APACyear {2002}}%
}]{%
Leino2002}
\APACinsertmetastar {%
Leino2002}%
\begin{APACrefauthors}%
Leino, K\BPBI R\BPBI M.%
\BCBT {}\ \BBA {} Nelson, G.%
\end{APACrefauthors}%
\unskip\
\newblock
\APACrefYearMonthDay{2002}{sep}{}.
\newblock
{\BBOQ}\APACrefatitle {Data Abstraction and Information Hiding} {Data
  abstraction and information hiding}.{\BBCQ}
\newblock
\APACjournalVolNumPages{ACM Trans. Program. Lang. Syst.}{24}{5}{491–553}.
\newblock
\begin{APACrefURL} \url{https://doi.org/10.1145/570886.570888} \end{APACrefURL}
\newblock
\begin{APACrefDOI} \doi{10.1145/570886.570888} \end{APACrefDOI}
\PrintBackRefs{\CurrentBib}

\bibitem [\protect \citeauthoryear {%
Leino%
\ \BBA {} W{\"u}stholz%
}{%
Leino%
\ \BBA {} W{\"u}stholz%
}{%
{\protect \APACyear {2014}}%
}]{%
leino2014dafny}
\APACinsertmetastar {%
leino2014dafny}%
\begin{APACrefauthors}%
Leino, K\BPBI R\BPBI M.%
\BCBT {}\ \BBA {} W{\"u}stholz, V.%
\end{APACrefauthors}%
\unskip\
\newblock
\APACrefYearMonthDay{2014}{}{}.
\newblock
{\BBOQ}\APACrefatitle {The {Dafny} integrated development environment} {The
  {Dafny} integrated development environment}.{\BBCQ}
\newblock
\APACjournalVolNumPages{\rm arXiv preprint arXiv:1404.6602}{}{}{}.
\PrintBackRefs{\CurrentBib}

\bibitem [\protect \citeauthoryear {%
B.~Liskov%
\ \BBA {} Guttag%
}{%
B.~Liskov%
\ \BBA {} Guttag%
}{%
{\protect \APACyear {2000}}%
}]{%
LiskovGuttag2000}
\APACinsertmetastar {%
LiskovGuttag2000}%
\begin{APACrefauthors}%
Liskov, B.%
\BCBT {}\ \BBA {} Guttag, J.%
\end{APACrefauthors}%
\unskip\
\newblock
\APACrefYear{2000}.
\newblock
\APACrefbtitle {Program Development in Java: Abstraction, Specification, and
  Object-Oriented Design} {Program development in java: Abstraction,
  specification, and object-oriented design}\ (\PrintOrdinal{1st}\ \BEd).
\newblock
\APACaddressPublisher{USA}{Addison-Wesley Longman Publishing Co., Inc.}
\PrintBackRefs{\CurrentBib}

\bibitem [\protect \citeauthoryear {%
B.~Liskov%
, Snyder%
, Atkinson%
\BCBL {}\ \BBA {} Schaffert%
}{%
B.~Liskov%
\ \protect \BOthers {.}}{%
{\protect \APACyear {1977}}%
}]{%
liskov1977}
\APACinsertmetastar {%
liskov1977}%
\begin{APACrefauthors}%
Liskov, B.%
, Snyder, A.%
, Atkinson, R.%
\BCBL {}\ \BBA {} Schaffert, C.%
\end{APACrefauthors}%
\unskip\
\newblock
\APACrefYearMonthDay{1977}{aug}{}.
\newblock
{\BBOQ}\APACrefatitle {Abstraction Mechanisms in CLU} {Abstraction mechanisms
  in clu}.{\BBCQ}
\newblock
\APACjournalVolNumPages{Commun. ACM}{20}{8}{564–576}.
\newblock
\begin{APACrefURL}
  \url{https://doi-org.ezproxy.elib10.ub.unimaas.nl/10.1145/359763.359789}
  \end{APACrefURL}
\newblock
\begin{APACrefDOI} \doi{10.1145/359763.359789} \end{APACrefDOI}
\PrintBackRefs{\CurrentBib}

\bibitem [\protect \citeauthoryear {%
B\BPBI H.~Liskov%
\ \BBA {} Wing%
}{%
B\BPBI H.~Liskov%
\ \BBA {} Wing%
}{%
{\protect \APACyear {1994}}%
}]{%
liskov1994behavioral}
\APACinsertmetastar {%
liskov1994behavioral}%
\begin{APACrefauthors}%
Liskov, B\BPBI H.%
\BCBT {}\ \BBA {} Wing, J\BPBI M.%
\end{APACrefauthors}%
\unskip\
\newblock
\APACrefYearMonthDay{1994}{}{}.
\newblock
{\BBOQ}\APACrefatitle {A behavioral notion of subtyping} {A behavioral notion
  of subtyping}.{\BBCQ}
\newblock
\APACjournalVolNumPages{ACM Transactions on Programming Languages and Systems
  (TOPLAS)}{16}{}{1811--1841}.
\PrintBackRefs{\CurrentBib}

\bibitem [\protect \citeauthoryear {%
Liu%
}{%
Liu%
}{%
{\protect \APACyear {2007}}%
}]{%
liu2007integrating}
\APACinsertmetastar {%
liu2007integrating}%
\begin{APACrefauthors}%
Liu, S.%
\end{APACrefauthors}%
\unskip\
\newblock
\APACrefYearMonthDay{2007}{}{}.
\newblock
{\BBOQ}\APACrefatitle {Integrating specification-based review and testing for
  detecting errors in programs} {Integrating specification-based review and
  testing for detecting errors in programs}.{\BBCQ}
\newblock
\BIn{} \APACrefbtitle {International Conference on Formal Engineering Methods}
  {International conference on formal engineering methods}\ (\BPGS\ 136--150).
\PrintBackRefs{\CurrentBib}

\bibitem [\protect \citeauthoryear {%
Meyer%
}{%
Meyer%
}{%
{\protect \APACyear {1997}}%
}]{%
meyer1997construction}
\APACinsertmetastar {%
meyer1997construction}%
\begin{APACrefauthors}%
Meyer, B.%
\end{APACrefauthors}%
\unskip\
\newblock
\APACrefYear{1997}.
\newblock
\APACrefbtitle {Object-Oriented Software Construction} {Object-oriented
  software construction}\ (\PrintOrdinal{2nd}\ \BEd).
\newblock
\APACaddressPublisher{}{Prentice-Hall}.
\PrintBackRefs{\CurrentBib}

\bibitem [\protect \citeauthoryear {%
Meyer%
}{%
Meyer%
}{%
{\protect \APACyear {2002}}%
}]{%
meyer2002design}
\APACinsertmetastar {%
meyer2002design}%
\begin{APACrefauthors}%
Meyer, B.%
\end{APACrefauthors}%
\unskip\
\newblock
\APACrefYear{2002}.
\newblock
\APACrefbtitle {Design by Contract} {Design by contract}.
\newblock
\APACaddressPublisher{}{Prentice-Hall}.
\PrintBackRefs{\CurrentBib}

\bibitem [\protect \citeauthoryear {%
Morgan%
}{%
Morgan%
}{%
{\protect \APACyear {1990}}%
}]{%
morgan1990programming}
\APACinsertmetastar {%
morgan1990programming}%
\begin{APACrefauthors}%
Morgan, C.%
\end{APACrefauthors}%
\unskip\
\newblock
\APACrefYear{1990}.
\newblock
\APACrefbtitle {Programming from Specifications} {Programming from
  specifications}.
\newblock
\APACaddressPublisher{}{Prentice-Hall}.
\PrintBackRefs{\CurrentBib}

\bibitem [\protect \citeauthoryear {%
Morris%
}{%
Morris%
}{%
{\protect \APACyear {1989}}%
}]{%
morris1989laws}
\APACinsertmetastar {%
morris1989laws}%
\begin{APACrefauthors}%
Morris, J\BPBI M.%
\end{APACrefauthors}%
\unskip\
\newblock
\APACrefYearMonthDay{1989}{}{}.
\newblock
{\BBOQ}\APACrefatitle {Laws of data refinement} {Laws of data
  refinement}.{\BBCQ}
\newblock
\APACjournalVolNumPages{Acta Informatica}{26}{4}{287--308}.
\PrintBackRefs{\CurrentBib}

\bibitem [\protect \citeauthoryear {%
Paige%
\ \BBA {} Ostroff%
}{%
Paige%
\ \BBA {} Ostroff%
}{%
{\protect \APACyear {2004}}%
}]{%
paige2004specification}
\APACinsertmetastar {%
paige2004specification}%
\begin{APACrefauthors}%
Paige, R\BPBI F.%
\BCBT {}\ \BBA {} Ostroff, J\BPBI S.%
\end{APACrefauthors}%
\unskip\
\newblock
\APACrefYearMonthDay{2004}{}{}.
\newblock
{\BBOQ}\APACrefatitle {Specification-driven design with {Eiffel} and agents for
  teaching lightweight formal methods} {Specification-driven design with
  {Eiffel} and agents for teaching lightweight formal methods}.{\BBCQ}
\newblock
\BIn{} \APACrefbtitle {International Conference on Technical Formal Methods}
  {International conference on technical formal methods}\ (\BPGS\ 107--123).
\PrintBackRefs{\CurrentBib}

\bibitem [\protect \citeauthoryear {%
Parnas%
}{%
Parnas%
}{%
{\protect \APACyear {2011}}%
}]{%
parnas2011precise}
\APACinsertmetastar {%
parnas2011precise}%
\begin{APACrefauthors}%
Parnas, D\BPBI L.%
\end{APACrefauthors}%
\unskip\
\newblock
\APACrefYearMonthDay{2011}{}{}.
\newblock
{\BBOQ}\APACrefatitle {Precise documentation: The key to better software}
  {Precise documentation: The key to better software}.{\BBCQ}
\newblock
\BIn{} \APACrefbtitle {The Future of Software Engineering} {The future of
  software engineering}\ (\BPGS\ 125--148).
\newblock
\APACaddressPublisher{}{Springer}.
\PrintBackRefs{\CurrentBib}

\bibitem [\protect \citeauthoryear {%
Passier%
, Bijlsma%
\BCBL {}\ \BBA {} Kuiper%
}{%
Passier%
\ \protect \BOthers {.}}{%
{\protect \APACyear {2022}}%
}]{%
passier2022}
\APACinsertmetastar {%
passier2022}%
\begin{APACrefauthors}%
Passier, H.%
, Bijlsma, L.%
\BCBL {}\ \BBA {} Kuiper, R.%
\end{APACrefauthors}%
\unskip\
\newblock
\APACrefYearMonthDay{2022}{}{}.
\newblock
\APACrefbtitle {Specification based {OO}-development: Procedural Guidance}
  {Specification based {OO}-development: Procedural guidance}\
  \APACbVolEdTR{}{\BTR{}\ \BNUM\ OUNL-CS-2022-6}.
\newblock
\APACaddressInstitution{Heerlen, The Netherlands}{Open universiteit}.
\newblock
\APACrefnote{Online at
  https://research.ou.nl/en/publications/specification-based-oo-development-procedural-guidance}
\PrintBackRefs{\CurrentBib}

\bibitem [\protect \citeauthoryear {%
Perrenet%
, Groote%
\BCBL {}\ \BBA {} Kaasenbrood%
}{%
Perrenet%
\ \protect \BOthers {.}}{%
{\protect \APACyear {2005}}%
}]{%
perrenet2005exploring}
\APACinsertmetastar {%
perrenet2005exploring}%
\begin{APACrefauthors}%
Perrenet, J.%
, Groote, J\BPBI F.%
\BCBL {}\ \BBA {} Kaasenbrood, E.%
\end{APACrefauthors}%
\unskip\
\newblock
\APACrefYearMonthDay{2005}{}{}.
\newblock
{\BBOQ}\APACrefatitle {Exploring students' understanding of the concept of
  algorithm: levels of abstraction} {Exploring students' understanding of the
  concept of algorithm: levels of abstraction}.{\BBCQ}
\newblock
\APACjournalVolNumPages{ACM SIGCSE Bulletin}{37}{3}{64--68}.
\PrintBackRefs{\CurrentBib}

\bibitem [\protect \citeauthoryear {%
Platzer%
}{%
Platzer%
}{%
{\protect \APACyear {2013}}%
}]{%
platzer2013teaching}
\APACinsertmetastar {%
platzer2013teaching}%
\begin{APACrefauthors}%
Platzer, A.%
\end{APACrefauthors}%
\unskip\
\newblock
\APACrefYearMonthDay{2013}{}{}.
\newblock
{\BBOQ}\APACrefatitle {Teaching CPS foundations with contracts} {Teaching cps
  foundations with contracts}.{\BBCQ}
\newblock
\APACjournalVolNumPages{\rm Carnegie Mellon University}{}{}{}.
\newblock
\begin{APACrefURL} \url{https://doi.org/10.1184/R1/6610286.v1} \end{APACrefURL}
\PrintBackRefs{\CurrentBib}

\bibitem [\protect \citeauthoryear {%
Polikarpova%
, Furia%
, Pei%
, Wei%
\BCBL {}\ \BBA {} Meyer%
}{%
Polikarpova%
\ \protect \BOthers {.}}{%
{\protect \APACyear {2013}}%
}]{%
polikarpova2013good}
\APACinsertmetastar {%
polikarpova2013good}%
\begin{APACrefauthors}%
Polikarpova, N.%
, Furia, C\BPBI A.%
, Pei, Y.%
, Wei, Y.%
\BCBL {}\ \BBA {} Meyer, B.%
\end{APACrefauthors}%
\unskip\
\newblock
\APACrefYearMonthDay{2013}{}{}.
\newblock
{\BBOQ}\APACrefatitle {What good are strong specifications?} {What good are
  strong specifications?}{\BBCQ}
\newblock
\BIn{} \APACrefbtitle {2013 35th International Conference on Software
  Engineering (ICSE)} {2013 35th international conference on software
  engineering (icse)}\ (\BPGS\ 262--271).
\PrintBackRefs{\CurrentBib}

\end{thebibliography}

\end{document}